\documentclass[twocolumn,preprintnumbers,amsmath,amssymb,superscriptaddress]{revtex4}
\usepackage{epsfig,bm}
\usepackage{ulem}
\usepackage{amsmath,color}

\begin{document}
\title{Examination of the $^{22}$C radius determination
    with interaction cross sections}
\author{T. Nagahisa}
\affiliation{Department of Physics,
  Hokkaido University, Sapporo 060-0810, Japan}
\author{W. Horiuchi}
\affiliation{Department of Physics,
  Hokkaido University, Sapporo 060-0810, Japan}

\begin{abstract}
A nuclear radius of $^{22}$C is investigated
with the total reaction cross sections at medium- to high-incident energies
in order to resolve the radius puzzle in which two recent interaction
  cross section measurements
using $^1$H and $^{12}$C targets show the quite different radii.
The cross sections of $^{22}$C are calculated consistently
for these target nuclei within a reliable microscopic framework,
the Glauber theory.
To describe appropriately such a reaction 
involving a spatially extended nucleus, 
the multiple scattering processes within the Glauber theory
are fully taken into account, that is,
the multi-dimensional integration in the Glauber amplitude
is evaluated using a Monte Carlo technique
without recourse to the optical-limit approximation.
We discuss the sensitivity of the spatially extended halo tail
to the total reaction cross sections.
The root-mean-square matter radius obtained in this study
is consistent with that extracted from
the recent cross section measurement on $^{12}$C target.
We show that the simultaneous reproduction
of the two recent measured cross sections is not feasible
within this framework.
\end{abstract}
\maketitle

\section{Introduction}

Advances in the radioactive ion beam facility
have revealed the exotic structure of short-lived neutron-rich unstable nuclei,
which has never been observed in stable nuclei,
such as halo structure~\cite{Tanihata85}.
The neutron dripline of carbon isotopes
is observed to be at $^{22}$C, which is known to be
the heaviest two-neutron halo nucleus has been found so far.
This nucleus is a key to the understanding of
the shell evolution along the neutron dripline
with the magicity at the neutron number 14 ($^{20}$C) and 16 ($^{22}$C).
The three-body $^{20}{\rm C}+n+n$ system is the so-called Borromean,
in which neither of the subsystems,
$^{20}{\rm C}$-$n$ and $n$-$n$, are bound,
leading to an extended two-neutron wave function
with $s$-wave dominance predicted
by the earlier three-body calculation~\cite{Horiuchi06}.
In fact, the two-neutron separation energy observed
is very small: 0.42$\pm$0.94\,MeV~\cite{Audi03}
and $0.14\pm 0.46$\,MeV~\cite{Gaudefroy12}.
The $s$-wave two-neutron halo structure
is further confirmed by the $^{20}$C fragment momentum distribution
measurement of the two neutron removal reaction
from $^{22}$C~\cite{Kobayashi12}.
This nucleus has attracted much attention 
not only to nuclear physics but also 
atomic physics in connection to
the Efimov physics~\cite{Yamashita11,Acharya13}.

A research interest has now been extended 
to reveal the exotic excitation mechanism
of $^{22}$C~\cite{Ershov12,Ogata13,Inakura14}.
However, the experimental situation on the $^{22}$C radius,
which is one of the most important and basic properties
of an atomic nucleus, has been still under discussion.
Since direct electron-scattering measurement
is not feasible at the moment, and a neutron radius is difficult to probe,
the nuclear radii of unstable nuclei have often been
studied by the total reaction or interaction cross sections at medium- 
and high-incident energies (several tens MeV to 1 GeV).
The first measurement of the interaction cross section
of $^{22}$C was performed in 2010 by Tanaka {\it et al.}~\cite{Tanaka10}.
The large interaction cross section on $^1$H target
incident at 40\,MeV/nucleon was measured 
1338$\pm 274$\,mb, resulting in a huge matter radius of 5.4$\pm$0.9\,fm
with large uncertainties.
Recently, high-precision measurement was made for
the interaction cross section on $^{12}$C target
incident at $\sim$240\,MeV/nucleon by Togano {\it et al.}~\cite{Togano16},
and the resultant radius is 3.44$\pm$0.08\,fm, which is quite
far from the previously extracted value 5.4$\pm$0.9\,fm~\cite{Tanaka10}.
Since the nuclear radius has often served as one of the
inputs to some theoretical models, e.g.,
Refs.~\cite{Yamashita11,Fortune12,Acharya13},
this demands appropriate reliable evaluation of the nuclear radius.

Here we focus on the theoretical investigation
of the nuclear radius of $^{22}$C
with the total reaction cross sections.
Use of such inclusive observables has some advantages:
The theory of describing the cross section is well established;
and the cross sections can be measured for almost all nuclei
as long as the beam intensity is sufficient;
and the different sensitivity to the nuclear density profile
can be controlled by a choice of a target nucleus and
an incident energy.
Systematic analyses of nuclear matter radii with
the total reaction cross sections on $^{12}$C target
incident at $\gtrsim 200$\,MeV/nucleon
have revealed structure changes and the role 
of excess neutrons of light neutron-rich unstable nuclei
~\cite{Takechi10, Minomo11,Minomo12,Sumi12,Horiuchi12,Watanabe14,Takechi14,Horiuchi15}.
We remark that the total reaction cross sections on $^1$H target
is also useful because the probe has different sensitivity
to protons and neutrons in the projectile nucleus depending on 
incident energies that can be used
to extract the neutron-skin thickness
of unstable nuclei~\cite{Horiuchi14,Horiuchi16}.

In this paper, we evaluate the nuclear radius of a two neutron halo nucleus,
$^{22}$C, from the total reaction cross sections 
on $^1$H and $^{12}$C targets, and
discuss the sensitivity of the halo tail to these cross sections.
We employ a reliable high-energy reaction theory,
the Glauber model~\cite{Glauber}, which is a microscopic
multiple-scattering theory starting
from the total nucleon-nucleon cross section.
In this work, the complete evaluation of the Glauber amplitude 
is made by using a Monte Carlo technique
in order to treat the extended two-neutron
halo wave function of $^{22}$C appropriately.
Also, we test the optical-limit-approximation (OLA),
a standard approximation of the Glauber model,
which has been used in many analyses of the radius extraction
(See Appendix for references),
and quantify the possible uncertainties with this approximation.

The paper is organized as follows.
Sec.~\ref{Glauber.sec} briefly explains
the Glauber model employed in this paper.
The Glauber amplitude which involves multi-dimensional integration
is introduced in this section.
Sec.~\ref{MonteCarlo.sec} is devoted to the evaluation
of the multi-dimensional integration using the Monte Carlo technique.
The explicit expression of the Glauber amplitude
is presented in Sec.~\ref{Glauberexp.sec}.
Sec.~\ref{wf.sec} explains
how to generate the wave function of $^{22}$C.
Monte Carlo configurations that crucially determine
the accuracy of the multi-dimensional integration
are generated in Sec.~\ref{MCconf.sec}.
They are tested in the total reaction cross section
calculations in Sec.~\ref{MCsig.sec}.
Our results are presented and discussed in Sec.~\ref{results.sec}.
A direct comparison between the theoretical and experimental
cross sections is made.
In Sec.~\ref{12C.sec}, the validity of our calculations
is confirmed with available experimental data
of $^{12}$C$+^{12}$C and $^{12}$C$+^1$H systems.
Then, we further confirm the reliability of our calculations
in the reactions involving $^{20}$C and $^{12}$C.
Sec.~\ref{22C.sec} presents our main results:
We describe the $^{22}$C$+^{12}$C and $^{22}$C$+^1$H reactions
in a consistent manner and discuss the possible uncertainties
in the radius extraction using the total reaction cross section.
The sensitivity of the halo tail of $^{22}$C to the total
reaction cross sections is also discussed.
The conclusion is drawn in Sec.~\ref{conclusion.sec}.
A detailed analysis of approximate treatment of the Glauber
amplitude is given in Appendix.

\section{Total reaction cross section in the Glauber model}
\label{Glauber.sec}

Here we consider a high-energy collision of
the projectile ($P$) and target ($T$) nuclei with
mass numbers $A_P$ and $A_T$, respectively.
The Glauber model~\cite{Glauber} is a microscopic multiple-scattering
theory which is widely used to study high-energy
nucleus-nucleus collisions.
With the help of the adiabatic and eikonal approximations,
the final state wave function of a projectile and target system,
$\Phi_f$, is greatly simplified as the product
of the ground-state wave functions of
the projectile $\Phi_0^P$ and the target $\Phi_0^T$ nuclei,
and the multiple-product of the phase-shift functions
of a nucleon-nucleon collision, $e^{i\chi_{NN}}$, as
\begin{align}
  \left|\Phi_f\right>=\exp
\left[i\sum_{j=1}^{A_P}\sum_{k=1}^{A_T}\chi_{NN}(\bm{b}+\hat{\bm{s}}_j^P-\hat{\bm{s}}_k^T)\right]
\left|\Phi_0^P\Phi_0^T\right>,
\end{align}
where $\bm{b}$ is the impact parameter vector 
perpendicular to the beam direction $z$,
and $\hat{\bm{s}}^{P}_j$ ($\hat{\bm{s}}^{T}_k$)
denotes the two-dimensional
single-particle coordinate operator
projected onto the $xy$-plane
of the $j$th ($k$th) nucleon
from the center-of-mass, abbreviated as cm,
of the projectile (target).

With this approximation, we only need
to evaluate the optical phase-shift function or the Glauber amplitude,
$e^{i\chi(\bm{b})}$, which includes all information of 
the elastic processes in the high-energy nuclear collision 
\begin{align}
e^{i\chi(\bm{b})}=\left<\Phi_0^P\Phi_0^T\right|
\prod_{j=1}^{A_P}\prod_{k=1}^{A_T}
\left[1-\Gamma_{NN}(\bm{b}+\hat{\bm{s}}_j^P-\hat{\bm{s}}_k^T)\right]
\left|\Phi_0^P\Phi_0^T\right>,
\label{optphase.eq}
\end{align}
where the profile function
$\Gamma_{NN}(\bm{b})=1-e^{i\chi_{NN}(\bm{b})}$ is introduced
for the sake of convenience.
The total reaction cross section is evaluated 
by integrating the reaction probability
\begin{align}
  P(\bm{b})=1-|e^{i\chi(\bm{b})}|^2,
\label{reacprob.eq}
\end{align}
over $\bm{b}$ as
\begin{align}
\sigma_R=\int d\bm{b}\, P(\bm{b}).
\label{sigr.eq}
\end{align}

The profile function is usually parametrized as~\cite{Ray79}
\begin{align}
  \Gamma_{NN}(\bm{b})=\frac{1-i\alpha_{NN}}{4\pi\beta_{NN}}
  \sigma_{NN}^{\rm tot}\exp\left[-\frac{\bm{b}^2}{2\beta_{NN}}\right],
\label{profile.eq}
\end{align}
where $\sigma_{NN}^{\rm tot}$, 
$\alpha_{NN}$, and $\beta_{NN}$
are the total nucleon-nucleon ($NN$) cross section,
the ratio between the real and imaginary parts
of the scattering amplitude at the forward angle,
and the so-called slope parameter, respectively.
Parameter sets for various incident energies
are listed in Ref.~\cite{Ibrahim08}
for proton-proton ($pp$) and proton-neutron ($pn$) 
are employed. The $nn$ ($np$) are taken to be the same as $pp$ ($pn$).
For the sake of simplicity,
hereafter we omit $NN$ in the profile function otherwise needed.
The validity of the parameter sets of the profile function
has already been confirmed
in a number of examples~\cite{Ibrahim09, Horiuchi10, Horiuchi12, Horiuchi14, Horiuchi16,Horiuchi17}.
The other inputs to the theory are
the wave functions of projectile and target nuclei.
Once these inputs are set, the theory has no adjustable parameter.
We do not consider the Coulomb breakup contributions
since the effects are negligible in systems
involving small $Z$ nuclei~\cite{Horiuchi10, Horiuchi16}.

\section{Evaluation of multi-dimensional integration in the Glauber amplitude}
\label{MonteCarlo.sec}

In general, the explicit evaluation of the Glauber amplitude
of Eq.~(\ref{optphase.eq}) is difficult because the expression involves
$3(A_P+A_T)$-dimensional integration.
For $^1$H target, it is possible to
reduce the dimension of the integral in the Glauber amplitude
when the projectile wave function
is represented by some specific forms such as
a Gaussian form~\cite{Ibrahim99} or a Slater determinant
of single-particle wave functions~\cite{Bassel68, Ibrahim09,Hatakeyama14,Hatakeyama15}.
For nucleus-nucleus scattering,
the explicit evaluation is in general tedious,
and thus one has to introduce
some approximations to reduce the complexity.
However, it is known that the standard optical-limit approximation
cannot be applied to nucleus-nucleus reactions involving
spatially extended nuclei, leading to
systematic uncertainties on the extraction
of the nuclear radii~\cite{Al-Khalili96,Al-Khalili96b}
(See also Appendix of this paper).
On the contrary, a Monte Carlo (MC) integration offers a direct way to
evaluate the multi-dimensional integration
in the Glauber amplitude
of Eq.~(\ref{optphase.eq})~\cite{Varga02, CPC03,Gibbs12}.
We take the same route
as the MC integration succeeds in its complete evaluation.

\subsection{Multi-dimensional integration in
  the Glauber amplitude}
\label{Glauberexp.sec}

The multi-dimensional integration
in Eq.~(\ref{optphase.eq}) is evaluated
using the MC integration.
For this purpose, we introduce the $A$-body density 
\begin{align}
\rho_{A}(\bar{\bm{r}}_1,\dots ,\bar{\bm{r}}_A)=
\left<\Phi_0\right|\prod_{i=1}^A\delta(\hat{\bar{\bm{r}}}_i-\bar{\bm{r}}_i)\left|\Phi_0\right>,
\label{manydens.eq}
\end{align}
where $\hat{\bar{\bm{r}}}_i$ is the single-particle coordinate operator of
the $i$th nucleon from the origin.
Then, the complete Glauber amplitude of Eq.~(\ref{optphase.eq}) reads 
\begin{align}
e^{i\chi(\bm{b})}&=\idotsint\, \left(\prod_{j=1}^{A_P}d\bar{\bm{r}}_j^P\right)
\left(\prod_{k=1}^{A_T}d\bar{\bm{r}}_k^T\right)\notag\\
&\times \rho_{A_P}^P(\bar{\bm{r}}_1^P,\dots ,\bar{\bm{r}}^P_{A_P})\rho_{A_T}^T(\bar{\bm{r}}_1^T,\dots ,\bar{\bm{r}}^T_{A_T})
\notag\\
&\times\prod_{j=1}^{A_P}\prod_{k=1}^{A_T}
\left[1-\Gamma(\bm{b}+\bm{s}_j^P-\bm{s}_k^T)\right],
\label{Glexplicit.eq}
\end{align}
where $\bm{s}_j^P$ ($\bm{s}_k^T$) denotes
the $xy$-component of the $j$th ($k$th) single-particle coordinate
from the cm coordinate of the projectile (target).
The product of the $A$-body densities of
the projectile and target nuclei, $\rho_{A_P}^{P}\rho_{A_T}^T$,
is the guiding function of the MC integration.
If appropriate MC configurations are given,
Eq.~(\ref{Glexplicit.eq}) can easily be evaluated
by summing up 
$\prod_{i=1}^{A_P}\prod_{j=1}^{A_T}[1-\Gamma(\bm{b}+\bm{s}_i^P-\bm{s}_j^T)]$
with these MC configurations at each $\bm{b}$.
Since the many-body operator,
$\prod_{i=1}^{A_P}\prod_{j=1}^{A_T}[1-\Gamma(\bm{b}+\hat{\bm{s}}_i^P-\hat{\bm{s}}_j^T)]$,
is translationally invariant, i.e.,
free from the cm motion,
the cm wave functions in $\Phi_0^P$ and $\Phi_0^T$ are
integrated out through the MC integration.
For spherical projectile and target nuclei,
the integration over $\bm{b}$ in Eq.~(\ref{sigr.eq}) is reduced
to one-dimensional one over $|\bm{b}|$
which is performed simply by the trapezoidal rule.

\subsection{Wave function}
\label{wf.sec}

The wave function is assumed 
to be the product of antisymmetrized 
neutron and proton wave functions 
\begin{align}
\Phi_0=\left(\mathcal{A}_n\Phi_n\right)
\left(\mathcal{A}_p\Phi_p\right)
\end{align}
with $\mathcal{A}_N$ being the antisymmetrizer
for proton ($N=p$) and neutron ($N=n$) defined by
\begin{align}
  \mathcal{A}_N=\frac{1}{\sqrt{\mathcal{N}_N!}}\sum_{(p_1, \dots, p_{\mathcal{N}_N})}^{\mathcal{N}_N!}
          {\rm sgn}(p_1, \dots, p_{\mathcal{N}_N})P_{(p_1,\dots, p_{\mathcal{N}_N})},
\end{align}
where the operator $P_{(p_1,\dots, p_{\mathcal{N}_N})}$ exchanges particle indices
and $\mathcal{N}_N$ denotes the number of proton or neutron.
For the sake of simplicity, we assume for $\Phi_N$ the product of 
the single-particle wave function $\phi_i(\bar{\bm{r}}_i)$
of the $i$th nucleon
\begin{align}
  \Phi_N=\prod_{i=1}^{\mathcal{N}_N}\phi_i(\bar{\bm{r}}_i).
\label{prodwf.eq}
\end{align}

In the present work, we have considered
the three nuclei, $^{12}$C, $^{20}$C, and $^{22}$C.
A configuration of the $^{12}$C wave function
is assumed to be $(0s_{1/2})^2(0p_{3/2})^4$ for both proton and neutron
with the harmonic-oscillator (HO) single-particle wave functions.
Since the charge radius of $^{12}$C is well known,
the HO length parameter can be fixed in such a way so as to
reproduce the point-proton radius, 2.33\,fm extracted from the
charge radius~\cite{Angeli13}.
For $^{20}$C and $^{22}$C,
single particle wave functions of $^{20}$C and
$^{22}$C systems are generated
from the phenomenological Woods-Saxon potential~\cite{BM,Horiuchi07}
\begin{align}
V(r)=-V_0f(r)+V_1(\hat{\bm{l}}\cdot\hat{\bm{s}})\frac{1}{r}\frac{d}{dr}f(r)+V_C(r),
\end{align}
where $f(r)=1/\left\{1+\exp\left[(r-R_N)/a\right]\right\}$
with $a=0.65$\,fm, $R_N=1.25A^{1/3}$\,fm.
$V_0$ is taken commonly for proton and neutron,
and $V_1=0.6875V_0$. $V_C$ is the Coulomb potential
with a uniform charge distribution with a sphere radius $R_N$,
which only acts on a proton.

We explain how we take the strength $V_0$ in the following:
A proton configuration is assumed to be $(0s_{1/2})^2(0p_{3/2})^4$.
The subshell closure of the neutron number $14$ and 16
is assumed for neutron configurations of $^{20}$C and $^{22}$C
and are taken respectively
as $(0s_{1/2})^2(0p_{3/2})^4(0p_{1/2})^2(0d_{5/2})^6$ for $^{20}$C
and $(0s_{1/2})^2(0p_{3/2})^4(0p_{1/2})^2(0d_{5/2})^6(1s_{1/2})^2$ for $^{22}$C.
These assumptions can be reasonable to describe $^{22}$C as
$^{20}$C$+n+n$ $s$-wave two-neutron halo structure~\cite{Horiuchi06}
which is confirmed by the $^{20}$C fragment momentum
distribution measurement of the two neutron removal reaction from
$^{22}$C~\cite{Kobayashi12}.
To simulate the two neutron halo structure of $^{22}$C, 
we firstly take $V_0$ commonly to all angular-momentum $l$ states
and fix it in such a way so as to reproduce
the interaction reaction cross section of $^{20}$C$+^{12}$C
measured at $\sim$900\,MeV~\cite{Ozawa01}.
Since a small $V_0$ value for $l=0$ ($V_0^{l=0}$) generates
the single-particle wave function with a long tail
that crucially determines the radius of the $^{22}$C,
we only vary $V_0^{l=0}$ as a free parameter
that controls the radius of $^{22}$C.

To perform the MC integration accurately,
we need to generate a large number of points,
typically $10^{6-8}$, which follow
the probability distribution $\rho_{A_P}^P\rho_{A_T}^T$ but
indeed it costs computational resources
because we have to take care of $({\mathcal{N}_p^{P}}!{\mathcal{N}_n^{P}}!{\mathcal{N}_p^T}!{\mathcal{N}_n^{T}}!)^2$
permutations for the projectile and target wave functions
coming from the bra and ket sides. 
In order to reduce the computational cost,
we consider to use the simple-product wave function
defined by Eq.~(\ref{prodwf.eq}).
Note that in the present case this assumption does not change
any one-body physical quantities such as nuclear radius and one-body density
but the $A$-body density of Eq.~(\ref{manydens.eq}) is modified
resulting in some cross section differences through
the Glauber amplitude of Eq.~(\ref{Glexplicit.eq}).
We confirm that the difference in the total reaction cross sections
on $^{1}$H target with the fully-antisymmetrized
and the simple-product wave functions
for $^{20}$C is small typically less than $\sim$1\%.
Therefore, for the practical reason,
we employ the simple-product wave functions
of $^{20}$C and $^{22}$C as Eq.~(\ref{prodwf.eq}).

\subsection{Monte Carlo configurations and nuclear radius}
\label{MCconf.sec}

The guiding function of the MC integration,
the $A$-body density~(\ref{manydens.eq}), is constructed
by a random walk with the Metropolis algorithm~\cite{Metropolis53}.
The number of spatial points (MC configurations) represented in
Cartesian coordinate $(x_1, y_1, z_1, \dots, x_A, y_A, z_A)$
are generated by the random walk with the step size $\Delta$.
The resulting MC configurations must follow
the probability distribution or the guiding function.
They are used to perform the multi-dimensional integration over
projectile and target coordinates.
The accuracy of the MC integration crucially depends
on the number of MC configurations $M$ and a choice of $\Delta$.
Since the total reaction cross section
is closely related to the nuclear size,
the MC configurations used in this paper
are required to reproduce at least
the root-mean-square matter radius (rms radius) of the $^{22}$C defined by
\begin{align}
  \sqrt{\left<\bar{r}^2\right>}=\sqrt{\frac{1}{A}
  \sum_{i=1}^A\int d\bar{\bm{r}}\,|\bar{\bm{r}}|^2|\phi_{i}(\bar{\bm{r}})|^2}
\label{rmsuc.eq}
\end{align}
with $A=22$.
We remark that the above expression (\ref{rmsuc.eq})
involves the cm contribution.
Though we exactly exclude the cm contribution through
the MC integration later, this uncorrected radius can be
used for a purpose to evaluate the precision of the MC integration.
As the wave function assumed in this paper is defined by
the product of the single-particle wave functions,
the integration becomes simple which can also be
evaluated accurately by a standard integration method,
the trapezoidal rule.

To optimize $\Delta$, we generate several probability distributions
for $^{22}$C with different $\Delta$ values and
calculate the rms radii defined in Eq.~(\ref{rmsuc.eq})
by the MC integration.
Then, they are compared with the ``exact'' rms radii
evaluated with the direct integration
in Eq.~(\ref{rmsuc.eq}) by the trapezoidal rule.
Finally, we set $\Delta=1.0$\,fm that minimizes the rms deviations of
the rms radii of $^{22}$C evaluated with the exact and the MC integration
ranging from $\sim 3$ to 4\,fm.
We note that in such extended wave functions
the optimal $\Delta$ value is larger than
that for a typical wave function.
In fact, $\Delta=0.25$\,fm is used as the optimal value
for $^{12}$C whose wave function is not much extended.

\begin{figure}[th]
  \begin{center}
    \epsfig{file=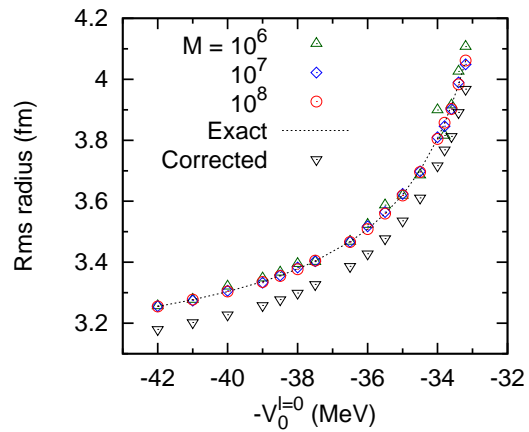, scale=1.3}
    \caption{
      Rms radii of $^{22}$C as a function of potential strengths
      for $l=0$ with different number of Monte Carlo configurations, $M$.
      Exact calculations with the center-of-mass (cm) contribution
      are plotted for comparison.
      The cm-free rms radii with $M=10^8$ (Corrected) are also plotted
      for comparison.
    See text for details.}
    \label{rms22C.fig}
  \end{center}
\end{figure}

Figure~\ref{rms22C.fig} displays
the cm uncorrected rms radii of $^{22}$C as a function of the potential
strengths $-V_0^{l=0}$ with different number of the MC configurations.
The exact rms radii 
are also plotted for comparison.
We confirm that desired MC configurations are
successfully generated with an appropriate choice of $\Delta$,
that is, all the MC configurations with $M=10^{6-8}$
reproduce perfectly the exact rms radii.
We will make further tests of these MC configurations 
for the multi-dimensional integration in the Glauber amplitude
in the next subsection.
The cm corrected rms radii of $^{22}$C
are also plotted in Fig.~\ref{rms22C.fig} with $M=10^8$,
which can be obtained by evaluating
the multi-dimensional integration 
\begin{align}
  \sqrt{\left<r^2\right>}=\sqrt{\frac{1}{A}\sum_{i=1}^{A}
    \idotsint\,\prod_{j=1}^Ad\bar{\bm{r}}_j\,
      |\bm{r}_i|^2\rho_A(\bar{\bm{r}}_1,\dots,\bar{\bm{r}}_A)}.
\label{rmscr.eq}
\end{align}
Taking $\rho_A(\bar{\bm{r}}_1,\dots,\bar{\bm{r}}_{A})$ as the guiding function,
one can easily perform the multi-dimensional integration
by summing up $|\bm{r}_i|^2$ ($\bm{r}_i=\bar{\bm{r}}_i-\bm{X}$
with $\bm{X}=\frac{1}{A}\sum_{i=1}^{A}\bar{\bm{r}}_i$)
using a set of the MC configurations.
The difference between the cm corrected and uncorrected
radii appears to be large typically $\sim 0.1$\,fm,
which cannot be neglected for the realistic calculations.

\subsection{Tests of Monte Carlo configurations in the total reaction cross section calculations}
\label{MCsig.sec}

Here we test the accuracy of the MC integration
in the total reaction cross section
calculations with respect to the number of the MC configurations.
For $^1$H target, when the projectile wave function
is represented by the product of the single-particle wave functions,
we can factorize the expression
and evaluate the complete Glauber amplitude without
recourse to the MC integration as~\cite{Bassel68, Ibrahim09} 
\begin{align}
e^{i\bar{\chi}(\bm{b})}&=
\left<\Phi_0\right|\prod_{j=1}^{A}
[1-\Gamma(\bm{b}+\hat{\bar{\bm{s}}}_j)]
\left|\Phi_0\right>\\
&=\idotsint\,\left(\prod_{j=1}^Ad\bar{\bm{r}}_j\right)\notag\\
&\times\rho_A(\bar{\bm{r}}_1,\dots,\bar{\bm{r}}_A)
\prod_{j=1}^{A}[1-\Gamma(\bm{b}+\bar{\bm{s}}_j)]
\label{Glp-nucl.eq}\\
&=\prod_{j=1}^{A}\left[1-\int d\bar{\bm{r}}\,\phi_j^*(\bar{\bm{r}})
\Gamma(\bm{b}+\bar{\bm{s}})\phi_j(\bar{\bm{r}})\right].
\label{Glp-nucl-fact.eq}
\end{align}
Eq.~(\ref{Glp-nucl.eq}) is the explicit form
for the MC integration,
while in Eq.~(\ref{Glp-nucl-fact.eq})
one can simply use the trapezoidal rule for
the integration over $\bm{\bar{r}}$.
Obviously, the above Glauber amplitude includes
the cm contribution but the expression is useful
for a test of the MC integration
as was done in the previous subsection.

\begin{figure}[th]
  \begin{center}
    \epsfig{file=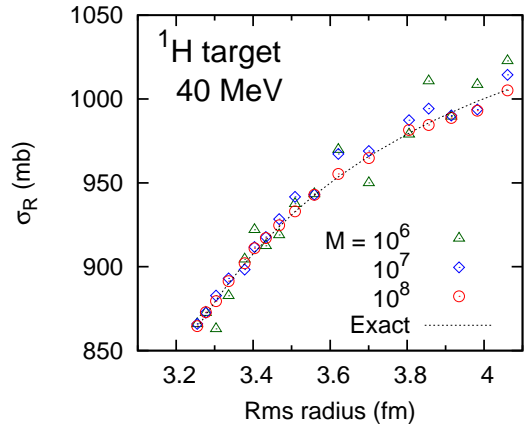, scale=1.3}
    \caption{Tests of total reaction cross sections
      of $^{22}$C on $^1$H target incident at 40\,MeV
      as a function of the cm uncorrected rms radii
      with different number of the MC configurations. 
      ``Exact'' values are also plotted for comparison.
      See text for details.}
    \label{rcstest22C-p.fig}
  \end{center}
\end{figure}

The incident energies are chosen as 40\,MeV and 240\,MeV
for $^1$H and $^{12}$C targets, respectively,
where the experimental data are available.
Here the incident energy is measured in MeV per nucleon
and for simplicity is written in MeV throughout this paper.
Figure~\ref{rcstest22C-p.fig} compares
the total reaction cross sections on $^{1}$H target
evaluated with different numbers of the MC configurations
as a function of the cm uncorrected rms radii.
In order to make a direct comparison with
the expression of Eqs.~(\ref{Glp-nucl.eq})
and ~(\ref{Glp-nucl-fact.eq}),
they are respectively evaluated by the MC and
trapezoidal (Exact) integration.
Though all the wave functions give almost the same 
rms radius as shown in Fig.~\ref{rms22C.fig},
the cross sections shows somewhat scattered
distributions, depending on the number of the
MC configurations, with $M=10^{6}$ and 10$^7$.
The cross sections converge to the exact values with increasing
the number of the MC configurations.
The deviations become at most by $\sim 1$\% with $M=10^8$.
We note that the convergence of the cross section is much
slower than that of an ordinary nuclear system, e.g., $^{12}$C and $^{20}$C
which typically need $M=10^6$ and 10$^7$, respectively.
More MC configurations are needed to have sufficient statistics
in the tail regions of the extended wave function of $^{22}$C.
In order to ensure the accuracy of
the total reaction cross sections of $^{22}$C on $^1$H target
within 1\% level,
we employ $M=10^8$ configurations for the MC integration.

\begin{figure}[th]
  \begin{center}
    \epsfig{file=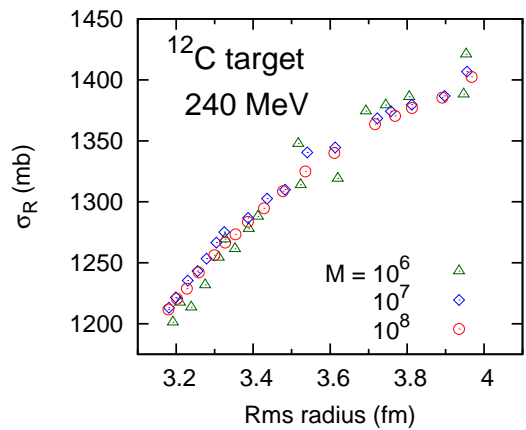, scale=1.3}
    \caption{Total reaction cross sections
      of $^{22}$C on $^{12}$C target incident at 240\,MeV
      as a function of the center-of-mass (cm)
      corrected rms radii with different number
      of the MC configurations. The cm contribution
      is exactly excluded in the calculations.}
    \label{rcstest22C-C.fig}
  \end{center}
\end{figure}

Next, we apply these MC configurations to the $^{22}$C$+^{12}$C case
where the factorization method of Eq.~(\ref{Glp-nucl-fact.eq})
can no longer be applied.
Figure~\ref{rcstest22C-C.fig} displays
the total reaction cross sections of
$^{22}$C on $^{12}$C target as a function of the rms radii.
The cm contribution is exactly removed through the MC integration
in Eq.~(\ref{rmscr.eq}).
The trend of the cross sections with respect to $M$
is similar to those on $^1$H target:
The cross section distributions are scattered with $M=10^{6}$ and 10$^7$
and a monotonic and smooth increase of the cross sections is obtained
with $M=10^8$ even at large rms radii.
We confirm that one can safely use the MC configurations with $M=10^8$ 
for the multi-dimensional integration in
the Glauber amplitude involving the very-extended $^{22}$C wave function
for the analysis of the total reaction cross sections
on both $^1$H and $^{12}$C targets.

\section{Results and discussions}
\label{results.sec}

\subsection{Comparison with measured cross sections of $^{12}$C and $^{20}$C}
\label{12C.sec}

\begin{figure}[th]
  \begin{center}
    \epsfig{file=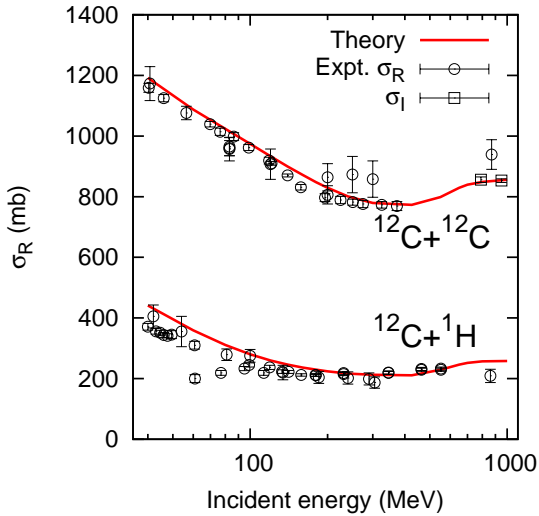, scale=1.3}
    \caption{Total reaction cross sections of $^{12}$C$+^{12}$C
      and $^{12}$C$+^{1}$H collisions as a function of incident energies.
      Experimental data of the total reaction ($\sigma_R$)
      and interaction ($\sigma_I)$ cross sections
      are taken from
      Refs.~\cite{takechi,perrin,zhang,fang00,kox,zheng,hostachy,jaros,Ozawa01b}
      for $^{12}$C$+^{12}$C and
      Refs.~\cite{Carlson96, Auce05} for $^{12}$C$+^{1}$H.
    }
    \label{rcs12C.fig}
  \end{center}
\end{figure}

Thus far, we have established that the accuracy of the MC integration
in the Glauber amplitude. In this subsection, we show the reliability
of our approach in comparison with available experimental cross section
data of $^{12}$C and $^{20}$C on $^{12}$C and $^1$H targets.

Figure~\ref{rcs12C.fig} displays
the total reaction cross sections on $^{12}$C and
$^1$H targets as a function of incident energies.
Our theory nicely reproduces the cross section data
at the low- to high-incident energies for both $^{12}$C and $^1$H targets.
The medium- to high-energy nuclear breakup processes
are described systematically very well.
Though the experimental data are scattered,
we see, at a close look, some deviations
from the experimental data with $^1$H target
below $\sim$100\,MeV and above $\sim 900$\,MeV
from the experimental values at most by 10\%.

\begin{figure}[th]
  \begin{center}
    \epsfig{file=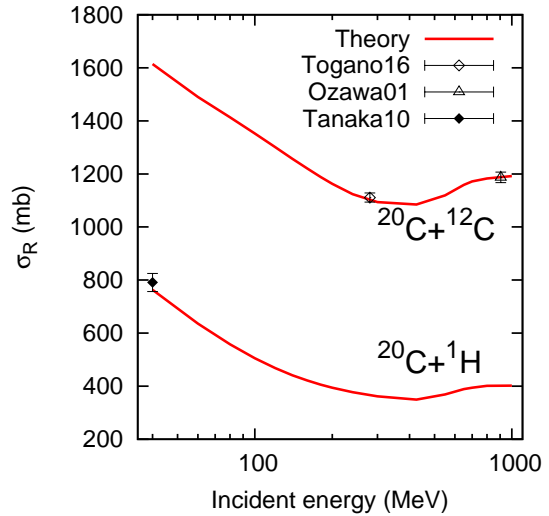, scale=1.3}
    \caption{Same as Fig.~\ref{rcs12C.fig}
      but of $^{20}$C.
      Experimental interaction cross section
      data are taken from Refs.~\cite{Ozawa01,Tanaka10,Togano16}.}
    \label{rcs20C.fig}
  \end{center}
\end{figure}

Figure~\ref{rcs20C.fig} plots the energy dependence of
the total reaction cross sections of $^{20}$C
on $^{12}$C and $^{1}$H targets.
The rms radius of $^{20}$C is 3.03\,fm which is determined
so as to reproduce the interaction cross section measured
at 905\,MeV~\cite{Ozawa01}.
We confirm that our calculations
are consistent with the interaction cross section data at 240\,MeV
on $^{12}$C target ~\cite{Togano16}
as well as that at 40\,MeV on $^{1}$H target~\cite{Tanaka10}.

\subsection{$^{22}$C: Nuclear radius vs total reaction cross sections}
\label{22C.sec}

\begin{figure}[th]
  \begin{center}
    \epsfig{file=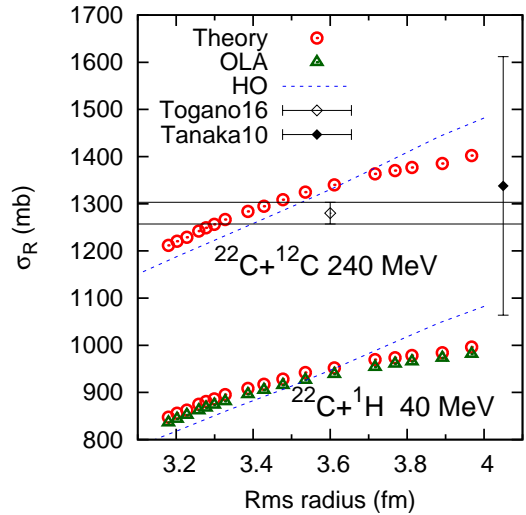, scale=1.3}
    \caption{Total reaction cross sections of $^{22}$C
      on $^{12}$C and $^{1}$H incident at 240\,MeV and 40\,MeV, respectively,
      as a function of rms radii of $^{22}$C.
      Thin lines denote the cross sections with the harmonic-oscillator (HO)
      wave function. See text for details.
      Experimental data are taken from Refs.~\cite{Tanaka10,Togano16}.
    }
    \label{rcs22C-exp.fig}
  \end{center}
\end{figure}

We have shown that
our theoretical model successfully describes
the total reaction cross sections involving
stable $^{12}$C and neutron-rich $^{20}$C at wide incident energies
for both $^{12}$C and $^{1}$H targets in a consistent manner.
Finally, let us discuss the controversy in the radius of $^{22}$C.
Figure~\ref{rcs22C-exp.fig} displays the total reaction
cross sections of $^{22}$C on $^{12}$C target incident at 240\,MeV
and on $^{1}$H target incident at 40\,MeV, respectively,
where the experimental data are available,
as a function of the rms radius.
The cross section data by Togano {\it et al.}~\cite{Togano16}
with uncertainties is indicated between two horizontal lines from which
we can extract the rms radius of $^{22}$C.
The resultant rms radius is 3.38$\pm$0.10\,fm
which is consistent with that
extracted by Togano {\it et al.}
using the sophisticated four-body Glauber model~\cite{Kucuk14},
3.44$\pm$0.08~\cite{Togano16}.
However, we find simultaneous reproduction of the cross section
data by Tanaka {\it et al.}~\cite{Tanaka10} is not possible
within $1\sigma$, that is, for $^1$H target, the experimental data is
far from the theoretical values
(However, it is consistent with $2\sigma$ as mentioned
  in Ref.~\cite{Togano16}).
Since our calculation is not feasible
for very large rms radius beyond $\sim$4\,fm,
we extrapolate the rms radius with
a form of $a\log[b(R-c)]$, where $R=\sqrt{\left<r^2\right>}$,
in which $a, b, c$ are determined by the least-square method.
The extrapolated radius is huge $\gtrsim 5$\,fm
at the lower limit (1$\sigma$) of the experimental cross section,
and never reach the central value of the experimental data
1338\,mb~\cite{Tanaka10} with the extrapolated
function based on our theoretical cross sections.

We discuss the possible uncertainties in the theoretical calculations.
We calculate the total reaction cross section on $^1$H target
with the OLA which was employed in the analysis of Ref.~\cite{Tanaka10}.
The phase-shift function of the OLA is given
as the leading order of the cumulant expansion 
of the complete Glauber amplitude~\cite{Glauber,Suzuki03}
\begin{align}
  i\chi_{\rm OLA}(\bm{b})=-\sum_{N=p,n}\int d\bm{r}\,\rho_N(\bm{r})
\Gamma_{pN}(\bm{b}-\bm{s}),
\end{align}
where $\bm{r}=(\bm{s},z)$ with $\bm{s}$
being a two-dimensional vector perpendicular to $z$,
and the translationally-invariant one-body density of the projectile
\begin{align}
  \rho_{N}(\bm{r})=\sum_{i=1}^{\mathcal{N}_N}\left<\Phi_N\right|
  \delta(\hat{\bm{r}}_i-\bm{r})
  \left|\Phi_N\right>.
\label{obdens.eq}
\end{align}
where $\hat{\bm{r}}_i$ denotes the $i$th single-particle
coordinate operator measured from the cm of the system.
The cm contribution in the one-body density is
exactly removed through the MC integration.
It is noted that this is one of the advantages of the present approach. 
In general, the removal of the cm contribution
needs some efforts. Some approximate methods for the removal prescribed, e.g.,
in Refs.~\cite{Negele70,Horiuchi07} becomes worse
since the square overlap of the HO and
the halo wave functions of $^{22}$C becomes 0.82-0.85
in the present range of the rms radii,
while it is larger than 0.99 for a non-halo nucleus, $^{20}$C.

The calculated total reaction cross sections with the OLA 
are displayed in Fig.~\ref{rcs22C-exp.fig}.
Here we only plot the OLA results on $^{1}$H target. 
More detailed comparisons between the complete Glauber calculation and
the OLA approximation for nucleus-nucleus scattering
are drawn in Appendix.
The difference between the complete Glauber and the OLA cross sections
is small approximately $1$\%, being the situation unchanged.

One may also think that the incident energy of 40\,MeV
is too low in the Glauber calculation.
As shown in Figs.~\ref{rcs12C.fig} and \ref{rcs20C.fig},
the theory reproduces fairly well the total reaction cross section
of $^{20}$C on $^{1}$H target even at 40\,MeV.
Since any excited bound state of $^{22}$C has not been observed so far,
the total reaction and interaction cross sections are equal
for $^{1}$H target
and its difference is expected to be small for $^{12}$C target.
The Coulomb breakup effect is expected to be small. For instance,
the contribution is estimated less than 1\% in the case of a
one-neutron halo nucleus, $^{31}$Ne on $^{12}$C target~\cite{Horiuchi10}.
It becomes even smaller in the case of $^1$H target.
Considering the theoretical uncertainties discussed above,
we conclude that the simultaneous reproduction
of both the experimental cross sections on $^{12}$C and $^1$H
in Refs.~\cite{Togano16,Tanaka10} is not possible
within the error bar.

Let us discuss what is actually probed by the total reaction cross sections
on $^{12}$C and $^1$H targets at those specific incident energies.
The total reaction cross sections at medium- to high-incident energies
are closely related to the nuclear radii of
colliding nuclei, $\sigma_R\sim \pi (R_P+R_T)^2$,
where $R_P$ ($R_T$) is the nuclear radius of the projectile (target) nucleus.
In fact, Figure~\ref{rcs22C-exp.fig} shows good proportionality
of the cross sections on the rms radii and this enhancement
  is similar for $^{12}$C and $^1$H targets.
It is interesting to note that
this increase becomes moderate for large rms radii.
To confirm whether this effect is due to the halo structure
or not, we generate a ``standard'' nucleus by assuming
for the $^{22}$C wave function the product of
the HO single-particle wave functions.
The cross sections with the HO wave function
are plotted in Fig.~\ref{rcs22C-exp.fig}
as a function of the rms radii which are controlled by
the HO oscillator length parameter.
The cross section firmly increases as the rms radius increases
which is in contrast to the case with the halo wave function.

\begin{figure}[th]
  \begin{center}
    \epsfig{file=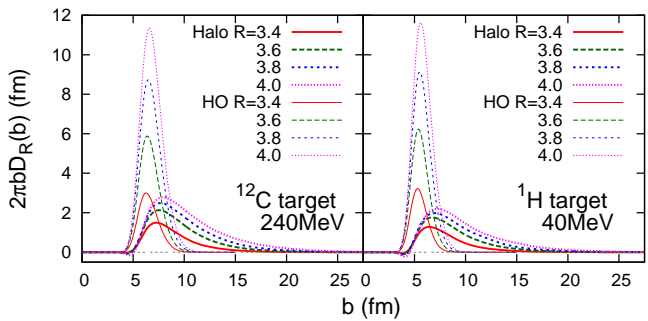, scale=0.95}
    \caption{Difference between two reaction probabilities
      of the $^{22}$C wave functions with different rms radii
on $^{12}$C target at 240\,MeV and on $^1$H target at 40\,MeV
      as a function of impact parameters.
      The reference reaction probability ($R=3.20$)
      is subtracted from each probabilities
      with $R=3.40$, 3.60, 3.80, and 4.00\,fm.
      See text for details.
      Thick lines denote the results with the halo wave functions,
      while thin lines denote those with the HO wave functions.}          
    \label{prob22C.fig}
  \end{center}
\end{figure}

In order to clarify the reasons
of the different cross section enhancement with
the halo and HO wave functions,
we show the evolution of the reaction probabilities
defined in Eq.~(\ref{reacprob.eq}) with respect
to the rms radius $R$.
For this purpose, we calculate the difference between two reaction probabilities
defined by
\begin{align}
D_R(\bm{b})=P(\bm{b})|_{R}-P(\bm{b})|_{R=3.20},
\end{align}
where the probability with the $^{22}$C wave function which gives
$R=3.20$\,fm is subtracted
to see clearly changes of the probabilities,
Figure~\ref{prob22C.fig} plots $D_R$
calculated with the halo and HO wave functions
as a function of the impact parameter $b=|\bm{b}|$. 
For both $^{12}$C and $^{1}$H targets,
the behavior of $D_R$ with the halo and HO wave functions
are quite different:
The enhancement of the reaction probability becomes smaller and smaller
with increasing the rms radius in the case of the halo wave function,
whereas $D_R$ increases monotonically
in the case of the HO wave function.
For the halo wave function, since this is very much extended,
only the weakly-bound two-neutron wave function is
contributed to the enhancement of $D_R$.
With large $R$, only dilute neutron tail contributes
to the nuclear radius but not much to the total reaction cross section,
leading to the moderate increase of the cross sections with large $R$
observed in Fig.~\ref{rcs22C-exp.fig}.
In the case of the HO wave function,
all nuclear orbits extend with increasing
the HO oscillator length that results in
the monotonic increase of the cross sections.

We note, however, the difference of the reaction probabilities
displayed in Fig.~\ref{prob22C.fig}
appears to be similar in both $^{12}$C and $^{1}$H targets.
This indicates that the sensitivity of the density profile of the projectile
does not depend much on the target nuclei, $^{12}$C and $^1$H,
for this set of the incident energies.
Since the $pn$ total cross section
as well as the range of the interaction become
large in such a low incident energy,
the contribution involving the two-neutron halo tail
becomes significant being comparable to the case of $^{12}$C target.
The fact is consistent with the discussion
given in Ref.~\cite{Ibrahim08} that showed the advantage of using
the low energy nuclear reaction with $^{1}$H target to probe the neutron
distribution, where the $pn$ total cross section
  becomes much larger than that of the $pp$ one.
  This can also be seen in comparison of the ordinary nucleus, $^{12}$C,
  and neutron-rich $^{20}$C reactions on $^1$H
  target displayed in Figs.~\ref{rcs12C.fig}
  and~\ref{rcs20C.fig}.

\begin{figure}[th]
  \begin{center}
    \epsfig{file=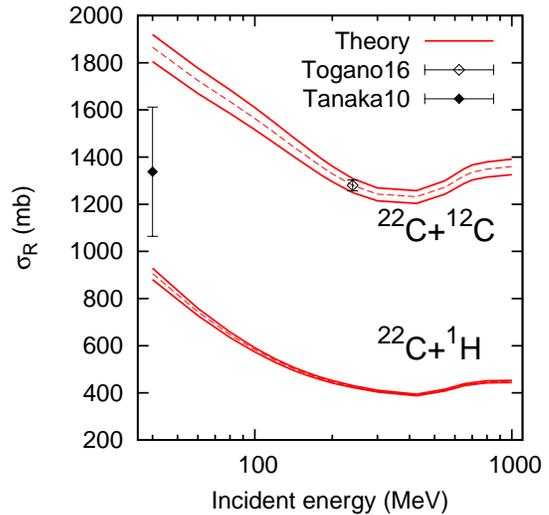, scale=1.3}
    \caption{Total reaction cross sections of $^{22}$C
      on $^{12}$C and $^1$H targets as a function
      of incident energies.
      The rms radius of $^{22}$C is set to be 3.38$\pm$0.10\,fm.
        The central value and the lower and upper bounds of the cross sections
        are indicated by dotted and solid lines, respectively.
    Experimental data are taken from Refs.~\cite{Tanaka10,Togano16}.}
    \label{rcs22C.fig}
  \end{center}
\end{figure}

Finally, we plot, in Fig.~\ref{rcs22C.fig},
the theoretical total reaction cross sections of $^{22}$C
as a function of the incident energies together
  with the available interaction cross section
  data~\cite{Tanaka10,Togano16}. We employ the wave function
giving $R=3.38\pm 0.10$\,fm taken consistently with
the recent interaction cross section data~\cite{Togano16}.
We again confirm that the target dependence is not large
at 40\,MeV for $^1$H target and at 240\,MeV for $^{12}$C target,
that is, the cross section variation
with respect to the radius change is almost the same.
The cross sections on $^{12}$C target
have some sensitivity of the halo tail at any incident energies,
whereas the ones on $^{1}$H target lose the sensitivity with increasing
the incident energy as the $pn$ total cross section becomes smaller.
In the figure, one can clearly see that the simultaneous reproduction
of the two experimental data within the error bar is not feasible.
Since we have only two experimental cross section data,
it is desired to have another data at different incident energy or target
in order to clarify that the $^{22}$C size is equivalent
to a radius of medium- ($A\sim 40$) or heavy- ($A\sim 200$) mass nuclei.
However, we already see theoretical consistency with the $^{20}$C
cross section data
for both $^1$H and $^{12}$C target in Fig.~\ref{rcs20C.fig}.
It is unlikely to have a huge radius $\gtrsim$5\,fm of $^{22}$C.

\section{Conclusion}
\label{conclusion.sec}

In order to resolve
the radius puzzle in $^{22}$C, we have investigated
the total reaction cross sections of $^{22}$C on $^{12}$C and $^1$H
targets incident at medium- to high-incident energies
within the framework of
a microscopic high-energy reaction theory, the Glauber model.
The complete optical phase-shift function or Glauber amplitude
in the Glauber model is evaluated with use of a Monte Carlo technique.

The calculated total reaction cross sections on $^{12}$C and $^{1}$H 
targets consistently reproduce the available experimental
cross section data for $^{12}$C and $^{20}$C.
We find that target dependence of the radius extraction of $^{22}$C
is small at $240$\,MeV for $^{12}$C target and 40\,MeV for $^{1}$H target.
We see, however, the simultaneous reproduction of the interaction
cross section data of $^{22}$C obtained by the two recent measurement
is not possible within the error bar ($1\sigma$). 
The root-mean-square (rms) matter radius of $^{22}$C
deduced from our analysis is consistent 
with the radius given in Ref.~\cite{Togano16}
using the interaction cross section incident on $^{12}$C target
at 240\,MeV, which corresponds to that of
an $A\sim 40$ nucleus.
We investigate possible uncertainties in
the theoretical model and they are actually small.
We conclude that it is unlikely to obtain
the huge rms matter radius of $\sim 5.4$ \,fm ($A\sim 200$)
shown in Ref.~\cite{Tanaka10}.

\acknowledgments

The authors thank J. Singh for careful reading of the manuscript.
This work was in part supported by JSPS KAKENHI Grant Numbers
  18K03635 and 18H04569.

\appendix

\section{Comparison with other approximations of the Glauber theory}

\begin{figure}[th]
  \begin{center}
    \epsfig{file=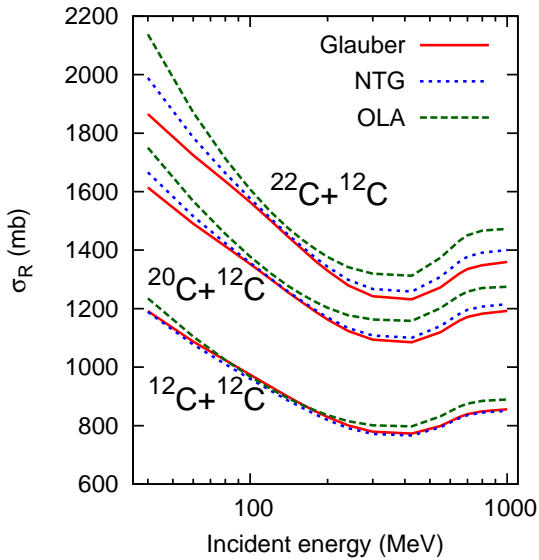, scale=1.3}
    \caption{Total reaction cross sections 
      of $^{12}$C, $^{20}$C, and $^{22}$C on $^{12}$C target
      as a function of incident energies calculated
with the complete Glauber amplitude, the NTG approximation, and the OLA.
See text for details.
The rms radii of $^{12}$C, $^{20}$C,
and $^{22}$C employed in the calculations
are 2.33, 3.03, and 3.38\,fm , respectively.}
    \label{Approx.fig}
  \end{center}
\end{figure}

In this appendix, we evaluate standard approximate methods
of the Glauber theory and quantify theoretical uncertainties
in nucleus-nucleus
total reaction cross section calculations.
In general, the evaluation of
the complete Glauber amplitude of Eq.~(\ref{optphase.eq})
requires tedious computations.
Therefore, the so-called optical-limit approximation (OLA)
has often been used as it only requires one-body density distributions
of the projectile and target nuclei.
This approximation relies on the cumulant expansion~\cite{Glauber,Suzuki03}
which offers series expansion in terms
of the fluctuation of the distribution function.
The expansion works well for such nuclei
having a standard density profile.
Contribution of the higher-order terms becomes more important
for an extended density distribution such as halo nuclei.
In fact, the standard OLA, which
only takes the leading term of the expansion, cannot
be applied to nucleus-nucleus reactions involving
halo nuclei as it leads to
some systematic uncertainties on the extraction
of the nuclear radii~\cite{Al-Khalili96,Al-Khalili96b}.

Though the OLA only takes the leading order of
the the consecutive product of the $NN$ phase-shift functions,
the approximation
already works well for the total reaction cross sections on $^1$H target
even they involve a halo nucleus as shown in
Refs.~\cite{Ibrahim99,Varga02} as well as in Fig.~\ref{rcs22C-exp.fig}
of the present paper.
The phase-shift function of the OLA is given
as the leading order of the cumulant expansion 
of the complete Glauber amplitude~\cite{Glauber,Suzuki03}
\begin{align}
  i\chi_{\rm OLA}(\bm{b})&=-\sum_{N,N^\prime=n,p}
  \iint \,d\bm{r}\,d\bm{r}^\prime\notag\\
&\times  \rho_N^P(\bm{r})\rho_{N^\prime}^T(\bm{r}^\prime)
\Gamma_{NN^\prime}(\bm{b}+\bm{s}-\bm{s}^\prime),
\label{OLA.eq}
\end{align}
where $\rho^{P}_N$ ($\rho^{T}_N$)
is the translationally-invariant one-body density of the projectile (target)
for proton $N=p$ and neutron $N=n$ defined in Eq.~(\ref{obdens.eq}).

For nucleus-nucleus scattering,
where the higher-order contribution would be sizable,
the Nucleon-Target formalism in the Glauber theory (NTG)~\cite{NTG},
has often been used:
\begin{align}
&i\chi_{\rm NTG}(\bm{b})
  =-\frac{1}{2}\sum_{N,N^\prime=n,p}\left\{\int\,d\bm{r}\rho_N^P(\bm{r})\right.\notag\\
  &\times \left[ 1+\exp \left(-\int\,d\bm{r}^\prime
    \rho_{N^\prime}^T(\bm{r}^\prime)
  \Gamma_{NN^\prime}(\bm{b}+\bm{s}-\bm{s}^\prime)\right)\right]\notag\\
&+\int\,d\bm{r}^\prime\rho_{N^\prime}^T(\bm{r}^\prime)\notag\\
&\times \left.\left[ 1+\exp \left(-\int\,d\bm{r} \rho_N^P(\bm{r})
\Gamma_{NN^\prime}(\bm{b}+\bm{s}^\prime-\bm{s})\right)\right]\right\}.
\label{NTG.eq}
\end{align}
Note that the same inputs of the OLA are required.
The NTG approximation has been applied to
a number of examples in the nucleus-nucleus
total reaction cross section calculations including stable
and neutron-rich isotopes~\cite{Horiuchi06,Horiuchi07,Takechi09,Ibrahim09,Horiuchi10,Hagino11,Kanungo11a,Kanungo11b,Horiuchi12,Horiuchi16,Horiuchi17,Urata17}.
Here we quantify the extent to which the higher-order terms
are included in the NTG approximation in comparison with the
complete Glauber calculation and the standard OLA.

Figure~\ref{Approx.fig}
plots the total reaction cross sections of $^{12}$C, $^{20}$C, and
$^{22}$C on $^{12}$C target as a function of the incident energies calculated
with the complete Glauber amplitude~(\ref{optphase.eq}),
the NTG approximation~(\ref{NTG.eq}), and the OLA~(\ref{OLA.eq}).
The wave functions of those nuclei are
taken consistently with the charge radius for $^{12}$C, and the interaction
cross sections at 900\,MeV~\cite{Ozawa01} for $^{20}$C
and at 240\,MeV~\cite{Togano16} for $^{22}$C. 
For $^{12}$C$+^{12}$C scattering, as already exemplified
in Refs.~\cite{Horiuchi07,Horiuchi12}, we again confirm that
the NTG gives better results than those obtained by the OLA
and takes care of most of the multiple-scattering effects missing in the OLA,
showing the cross sections much closer to the complete Glauber calculations.
The NTG approximation also works well for $^{20}$C
but large deviation appears with the OLA.
For $^{22}$C, as expected, the OLA considerably deviates
from the calculated cross sections obtained
with the complete Glauber amplitude.
The deviations of these approximations from the complete calculation
appear to be minimum at around 100-200\,MeV.
The NTG always gives better results than those of the OLA
but it is still not sufficient at low- and high-incident energies,
say 3\% deviation at 1000\,MeV from the complete calculation.
Though the deviations of these approximations are smaller at 240\,MeV,
these theoretical uncertainties actually affect the radius extraction
from the measured cross section data~\cite{Togano16}:
The extracted radii are $R=3.33\pm 0.09$ and $3.23\pm 0.07$\,fm
with the NTG and OLA, respectively,
while $R=3.38\pm 0.10$\,fm with the complete Glauber calculation.
The deviations become even larger with increasing the halo tail
and at different incident energies.
Here we have seen that the NTG approximation works well
for the standard density profile but not for the halo density.
One needs to care about the uncertainties included
in these approximations when the nuclear radius
is extracted from the total reaction cross section on $^{12}$C target.


\begin{thebibliography}{99}
\bibitem{Tanihata85} I. Tanihata, H. Hamagaki, O. Hashimoto, Y. Shida,
  N. Yoshikawa {\it et al.}, Phys. Rev. Lett. {\bf 55},  2676 (1985).
\bibitem{Horiuchi06} W. Horiuchi and Y. Suzuki,
  Phys. Rev. C {\bf 74} 034311 (2006).
\bibitem{Audi03}
  G. Audi, A.~H. Wapstra, and C. Thibault, Nucl. Phys. {\bf A 729},
  337 (2003). 
\bibitem{Gaudefroy12} L. Gaudefroy, W. Mittig, N.~A. Orr,
  S. Varet, M. Chartier {\it et al.},
  Phys. Rev. Lett. {\bf 109}, 202503 (2012).
\bibitem{Kobayashi12} N. Kobayashi, T. Nakamura, J.~A. Tostevin,
  Y. Kondo, N. Aoi {\it et al.},
  Phys. Rev. C {\bf 86}, 054604 (2012).
\bibitem{Yamashita11} M.~T. Yamashita,
  R.~S. Marques de Carvalho, T. Frederico, and L. Tomio,
Phys. Lett. {\bf B 697}, 90 (2011).
\bibitem{Acharya13}
B. Acharya, C. Ji, and D.~R. Phillips, Phys. Lett. {\bf B 723}, 196 (2013).
\bibitem{Ershov12} S.~N. Ershov, J.~S. Vaagen, and
  M.~V. Zhukov, Phys. Rev. C {\bf 86}, 034331 (2012).
\bibitem{Ogata13} K. Ogata, T. Myo, T. Furumoto, T. Matsumoto, and M. Yahiro,
  Phys. Rev. C {\bf 88}, 024616 (2013).
\bibitem{Inakura14} T. Inakura, W. Horiuchi, Y. Suzuki, and T. Nakatsukasa,
  Phys. Rev. C {\bf 89}, 064316 (2014).  
\bibitem{Tanaka10} K. Tanaka, T. Yamaguchi, T. Suzuki, T. Ohtsubo, M. Fukuda {\it et al.}, Phys. Rev. Lett. {\bf 104}, 062701 (2010). 
\bibitem{Togano16} Y. Togano, T. Nakamura, Y. Kondo, J.~A. Tostevin,
  A.~T. Saito {\it et al.}, Phys. Lett. {\bf B 761}, 412-418 (2016).
\bibitem{Fortune12} H.~T. Fortune and R. Sherr,
  Phys. Rev. C {\bf 85}, 027303 (2012).    
\bibitem{Takechi10} M. Takechi, T. Ohtsubo, T. Kuboki, M. Fukuda, D. Nishimura {\it et al.}, Mod. Phys. Lett. A {\bf 25}, 1878 (2010).
\bibitem{Minomo11} K. Minomo, T. Sumi, M. Kimura, K. Ogata, Y.~R. Shimizu,
  and M. Yahiro, Phys. Rev. C {\bf 84}, 034602 (2011).
\bibitem{Minomo12} K. Minomo, T. Sumi, M. Kimura, K. Ogata, Y.~R. Shimizu,
  and M. Yahiro, Phys. Rev. Lett. {\bf 108}, 052503 (2012).
\bibitem{Sumi12} T. Sumi, K. Minomo, S. Tagami, M. Kimura, T. Matsumoto
  {\it et al.}, Phys. Rev. C {\bf 85}, 
064613 (2012).
\bibitem{Horiuchi12} W. Horiuchi, T. Inakura, T. Nakatsukasa, and Y. Suzuki,
Phys. Rev. C {\bf 86}, 024614 (2012).
\bibitem{Watanabe14} S. Watanabe, K. Minomo, M. Shimada, S. Tagami, M. Kimura
  {\it et al.},
  Phys. Rev. C {\bf 89}, 044610 (2014).
\bibitem{Takechi14} M. Takechi, S. Suzuki, D. Nishimura, M. Fukuda, T. Ohtsubo {\it et al.}, Phys. Rev. C {\bf 90}, 061305(R) (2014).
\bibitem{Horiuchi15} W. Horiuchi, T. Inakura, T. Nakatsukasa, and Y. Suzuki,
  JPS Conf. Proc. {\bf 6}, 030079 (2015).  
\bibitem{Horiuchi14}
  W. Horiuchi, Y. Suzuki, and T. Inakura,
  Phys. Rev. C {\bf 89}, 011601 (R) (2014).
\bibitem{Horiuchi16}
  W. Horiuchi, S. Hatakeyama, S. Ebata, and Y. Suzuki,
  Phys. Rev. C {\bf 93}, 044611 (2016).
\bibitem{Glauber} R.~J. Glauber, {\it Lectures in Theoretical Physics}, edited by W.~E. Brittin and L.~G. Dunham (Interscience, New York, 1959), Vol. 1, p.315.
\bibitem{Ray79} L. Ray, Phys. Rev. C {\bf 20}, 1857 (1979).  
\bibitem{Ibrahim08} B. Abu-Ibrahim, W. Horiuchi, A. Kohama, and Y. Suzuki, 
                    Phys. Rev. C {\bf 77}, 034607 (2008);
                    {\it ibid} {\bf 80}, 029903 (2009); {\bf 81}, 019901 (2010).
\bibitem{Horiuchi10} W. Horiuchi, Y. Suzuki, P. Capel, and D. Baye,
  Phys. Rev. C {\bf 81}, 024606 (2010).
\bibitem{Ibrahim09} B. Abu-Ibrahim, S. Iwasaki, W. Horiuchi, A. Kohama, and Y. Suzuki, J. Phys. Soc. Jpn., Vol. 78, 044201 (2009).
\bibitem{Horiuchi17} W. Horiuchi, S. Hatakeyama, S. Ebata, and Y. Suzuki,
  Phys. Rev. C {\bf 96}, 024605 (2017).
\bibitem{Ibrahim99} B. Abu-Ibrahim, K. Fujimura, and Y. Suzuki,
  Nucl. Phys. {\bf A 657}, 391 (1999).
\bibitem{Bassel68} R.~H. Bassel and C. Wilkin, 
  Phys. Rev. {\bf 174}, 1179 (1968).
\bibitem{Hatakeyama14} S. Hatakeyama, S. Ebata, W. Horiuchi, and M. Kimura,
J. Phys.: Conf. Ser. {\bf 569}, 012050 (2014).
\bibitem{Hatakeyama15} S. Hatakeyama, S. Ebata, W. Horiuchi, and M. Kimura,
  JPS Conf. Proc., Vol. {\bf 6}, 030096 (2015).  
\bibitem{Al-Khalili96} J.~S. Al-Khalili and J.~A. Tostevin,
  Phys. Rev. Lett. {\bf 76}, 3903 (1996).
\bibitem{Al-Khalili96b} J.~S. Al-Khalili, J.~A. Tostevin,
and I.~J. Thompson,
  Phys. Rev. C {\bf 54}, 1843 (1996).
\bibitem{Varga02} K. Varga, S.~C. Pieper, Y. Suzuki, and R.~B. Wiringa,
  Phys. Rev. C {\bf 66}, 034611 (2002).
\bibitem{CPC03} B. Abu-Ibrahim, Y. Ogawa, Y. Suzuki, and I. Tanihata,
  Comp. Phys. Commun. {\bf 151}, 369 (2003).
\bibitem{Gibbs12} W.~R. Gibbs and J.~P. Dedonder,
  Phys. Rev. C {\bf 86}, 024604 (2012).
\bibitem{Angeli13} I. Angeli, K.~P. Marinova, At. Data. Nucl. Data Tables {\bf 99}, 69 (2013).
\bibitem{BM}
A. Bohr and B.~R. Mottelson, Nuclear Structure, Vol. I \& II (W.~A. Benjamin,
New York, 1975).
\bibitem{Horiuchi07} W. Horiuchi, Y. Suzuki, B. Abu-Ibrahim, and A. Kohama, Phys. Rev. C {\bf 75}, 044607 (2007).
\bibitem{Ozawa01} A. Ozawa, O. Bochkarev, L. Chulkov, D. Cortina, H. Geissel
  {\it et al.}, Nucl. Phys. {\bf A 691}, 599 (2001).
\bibitem{Metropolis53} N. Metropolis, A. Rosenbluth, M. Rosenbluth, E. Teller,
  J. Chem. Phys. {\bf 21}, 1087 (1953).
\bibitem{takechi} M. Takechi, M. Fukuda, M. Mihara, T. Chinda,
        T. Matsumasa {\it et al.}, Eur. Phys. J. A 25, s01, 217 (2005)
        and private communication.
\bibitem{perrin} C. Perrin, S. Kox, N. Longequeue, J.~B. Viano,
        M. Buenerd {\it et al.}, 
        Phys. Rev. Lett. {\bf 49}, 1905 (1982).
\bibitem{zhang} H.Y. Zhang, W.Q. Shen, Z.Z. Ren, Y.G. Ma,
        W.~Z. Jiang {\it et al.}, 
        Nucl. Phys. {\bf A 707}, 303 (2002).
\bibitem{fang00} D.Q. Fang, W.Q. Shen, J. Feng, X.~Z. Cai, 
        J.~S. Wang {\it et al.}, 
        Phys. Rev. C {\bf 61}, 064311 (2000).
\bibitem{kox} S. Kox, A. Gamp, C. Perrin, J. Arvieux, R. Bertholet 
        {\it et al.}, Phys. Rev. C {\bf 35}, 1678 (1987).
\bibitem{zheng} T. Zheng, T. Yamaguchi, A. Ozawa, M. Chiba, 
        R. Kanungo {\it et al.}, Nucl. Phys. {\bf A 709}, 103 (2002).
\bibitem{hostachy} J.~Y. Hostachy, M. Buenerd, J. Chauvin, D. Lebrun, 
        Ph. Martin {\it et al.}, Nucl. Phys. {\bf A 490}, 441 (1988).
\bibitem{jaros} J. Jaros, A. Wagner, L. Anderson, O. Chamberlain,
        R.~Z. Fuzesy {\it et al.}, 
        Phys. Rev. C {\bf 18}, 2273 (1978).
\bibitem{Ozawa01b} A. Ozawa, T. Suzuki, I. Tanihata,
        Nucl. Phys. {\bf A 693}, 32 (2001).
\bibitem{Carlson96} R.~F. Carlson, At. Dat. Nucl. Dat. Tables {\bf 63}, 93 (1996).
\bibitem{Auce05} A. Auce, A. Ingemarsson, R. Johansson, M. Lantz, G. Tibell
  {\it et al.}, Phys. Rev. C {\bf 71}, 064606 (2005).
\bibitem{Kucuk14} Y. Kucuk and J.~A. Tostevin, Phys. Rev C {\bf 89},
  034607 (2014).  
\bibitem{Suzuki03} Y. Suzuki, R.~G. Lovas, K. Yabana, and K. Varga,
  {\it Structure and reactions of light exotic nuclei} (Taylor \& Francis, London, 2003).
  \bibitem{Negele70} J.~W. Negele, Phys. Rev. C {\bf 1}, 1260 (1970).
  \bibitem{NTG} B. Abu-Ibrahim and Y. Suzuki, Phys. Rev. C {\bf 61}, 
  051601 (R) (2000).                               
\bibitem{Takechi09} M. Takechi, M. Fukuda, M. Mihara, K. Tanaka, T. Chinda {\it et al.}, Phys. Rev. C {\bf 79}, 061601(R) (2009).
\bibitem{Hagino11} K. Hagino and H. Sagawa,
  Phys. Rev. C {\bf 84}, 011303(R) (2011).
\bibitem{Kanungo11a} R. Kanungo, A. Prochazka, W. Horiuchi, C. Nociforo,
  T. Aumann {\it et al.}, Phys. Rev. C {\bf 83}, 021302(R) (2011).
\bibitem{Kanungo11b} R. Kanungo, A. Prochazka, M. Uchida, W. Horiuchi, G. Hagen
  {\it et al.}, Phys. Rev. C {\bf 84}, 061304(R) (2011).
\bibitem{Urata17} Y. Urata, K. Hagino, and H. Sagawa,
  Phys. Rev. C {\bf 96}, 064311 (2017).
\end{thebibliography}
\end{document}